\documentclass[sigconf]{acmart}
\usepackage{tabularx, tcolorbox, framed, longtable, stfloats, mathtools}

\AtBeginDocument{%
  }

\setcopyright{acmlicensed}
\copyrightyear{2018}
\acmYear{2018}
\acmDOI{XXXXXXX.XXXXXXX}
\acmConference[SBQS25]{Brazilian Symposium on Software Quality}{Nov 04--07, 2025}{São José dos Campos, SP}
\setcopyright{none}
\renewcommand\footnotetextcopyrightpermission[1]{}

\begin{document}
\renewcommand\abstractname{ABSTRACT}
\renewcommand\keywordsname{KEYWORDS}
\renewcommand\refname{REFERENCES}

\title[Will AI also replace inspectors? Investigating the potential of generative AIs in\newline usability inspection]{Will AI also replace inspectors? Investigating the potential of generative AIs in usability inspection}


\author{Luis F. G. Campos}
\orcid{0009-0004-3935-0734}
\affiliation{%
  \institution{Federal University of Technology - Paraná (UTFPR)}
  \city{Campo Mourão}
  \state{Paraná}
  \country{Brazil}
}
\email{luisfelipecampos@alunos.utfpr.edu.br}

\author{Leonardo C. Marques}
\orcid{0000-0002-3645-7606}
\affiliation{%
  \institution{SiDi - Intelligence \& Innovation Center}
  \city{Manaus}
  \state{Amazonas}
  \country{Brazil}
}
\email{lc.marques@sidi.org.br}

\author{Walter T. Nakamura}
\orcid{0000-0001-5451-3109}
\affiliation{%
  \institution{Federal University of Technology - Paraná (UTFPR)}
  \city{Campo Mourão}
  \state{Paraná}
  \country{Brazil}
}
\email{waltertakashi@utfpr.edu.br}

\renewcommand{\shortauthors}{Campos et al.}

\begin{abstract}
  Usability inspection is a well-established technique for identifying interaction issues in software interfaces, thereby contributing to improved product quality. However, it is a costly process that requires time and specialized knowledge from inspectors. With advances in Artificial Intelligence (AI), new opportunities have emerged to support this task, particularly through generative models capable of interpreting interfaces and performing inspections more efficiently. This study examines the performance of generative AIs in identifying usability problems, comparing them to those of experienced human inspectors. A software prototype was evaluated by four specialists and two AI models (GPT-4o and Gemini 2.5 Flash), using metrics such as precision, recall, and F1-score. While inspectors achieved the highest levels of precision and overall coverage, the AIs demonstrated high individual performance and discovered many novel defects, but with a higher rate of false positives and redundant reports. The combination of AIs and human inspectors produced the best results, revealing their complementarity. These findings suggest that AI, in its current stage, cannot replace human inspectors but can serve as a valuable augmentation tool to improve efficiency and expand defect coverage. The results provide evidence based on quantitative analysis to inform the discussion on the role of AI in usability inspections, pointing to viable paths for its complementary use in software quality assessment contexts.  
\end{abstract}


\maketitle

\section{Introduction}

The quality of a software system’s usability reflects how well it enables users to achieve their goals efficiently, effectively, and with satisfaction in a given context \cite{nasir2022usability}. Good usability is crucial in a voluntary use context, where insufficient support for task completion can lead users to stop using a software \cite{or2024inspection}.

Usability testing and usability inspection are among the most widely adopted methods for evaluating usability, each with distinct characteristics and limitations \cite{nasir2022usability}. While traditional usability testing often involves high costs, limited access to representative users, and logistical difficulties in replicating real-world usage scenarios, usability inspection relies on the judgment of experienced evaluators, called inspectors. The inspectors usually apply heuristic principles to identify potential usability problems, helping to uncover issues related to how effectively and efficiently users can complete tasks and their overall satisfaction with the system \cite{nielsen1994heuristic}. However, inspecting, documenting, distinguishing, analyzing, and synthesizing a summarized list of unique usability issues is time-consuming. Therefore, using Artificial Intelligence (AI) presents a promising alternative to support or even automate usability inspection activities.

AI has gained attention and interest and has been widely adopted in the work practices of usability practitioners \cite{abbas2022user}. One of the primary areas in which practitioners leverage AI as a support tool is usability testing and inspection \cite{fan2022human}. With the growing emergence of Generative Artificial Intelligence (GenAI) technologies and platforms such as ChatGPT \footnote{https://chat.openai.com/}, Gemini \footnote{https://gemini.google.com/}, and Copilot \footnote{https://copilot.microsoft.com/chats/}, a wide range of industries are increasingly exploring their adoption \cite{takaffoli2024generative}.

However, there is currently a lack of understanding and empirical evidence regarding whether GenAI technologies can perform usability inspections at a level comparable to those conducted by human inspectors. We argue that gaining this understanding is essential for the effective adoption of GenAI in inspection activities and for ensuring its responsible use in supporting the inspection process.

Therefore, in this work, we aim to compare the results of a usability inspection of a prototype conducted by GenAI with those performed by human evaluators. Our primary interest is to generate evidence-based results regarding the efficacy of the inspection results produced by ChatGPT-4o and Gemini 2.5 Flash and those produced by human usability inspectors. To do so, we conducted a usability inspection of an interface prototype with tasks defined for both AI models and human inspectors. The results showed that while AI demonstrated higher precision and completed the tasks faster, both AI and humans exhibited low recall. This suggests that combining AI and human expertise could improve overall problem detection, as neither approach alone identifies most usability issues.

The remainder of this paper is organized as follows: Section 2 contextualizes this research, providing background about usability inspection, GenAIs, and presenting related work. Section 3 presents the methodology we adopted in this study. Section 4 presents the results. Section 5 discusses the main findings. Section 6 presents threats to validity. Finally, section 7 concludes this paper by presenting an overview of the results and future work.

\section{Background}
This section outlines the key concepts related to our research and related work.

\subsection{Usability Inspection and Heuristic Evaluation}

Conducting usability inspections during the software project (or design) phase is an essential practice, as it enables the identification of usability issues and the formulation of targeted redesign suggestions to enhance software usability \cite{or2024inspection}. This type of evaluation can improve product performance, resulting in tangible and intangible benefits for end-users, developers, and stakeholders involved in the software development. These benefits are well-recognized and outlined below:

\begin{itemize}
    \item Usability inspection identifies errors that users may encounter when interacting with a product or system.
    \item One of the key advantages of usability inspection methods is that they are simple, fast to apply, and do not require fully functional software. This makes them especially valuable in the early stages of software development, when usability problems can be identified and addressed at a lower cost \cite{tao2012usability}.
    \item Before software deployment, usability inspections serve as a safeguard to ensure that performance and safety standards are met \cite{or2024inspection}.
    \item Ultimately, the primary goal of usability inspection is to reduce usability-related problems. A product with fewer usability issues will likely succeed in the market.
\end{itemize}

One of the primary applied usability inspection methods is Heuristic Evaluation. Heuristic evaluation is an analytical usability inspection method that involves assessing a software's User Interface (UI) against recognized usability principles, often referred to as heuristics \cite{molich1990improving}. The primary goal is to identify usability issues and determine whether the software aligns with established guidelines. Based on this analysis, evaluators generate redesign recommendations to improve the product's overall usability and effectiveness.

The effectiveness of a heuristic evaluation depends significantly on the evaluators' backgrounds \cite{molich1990improving}. Ideally, evaluators should possess usability expertise and be familiar with the product or domain. Research shows that usability specialists familiar with the system tend to identify more usability issues than those without such experience. While individual evaluators may detect only a portion of usability problems, studies suggest that involving three to five evaluators is sufficient to uncover most issues in small-scale systems \cite{virzi1990streamlining}. This approach balances evaluation coverage and resource efficiency, making it a practical method for early-stage design validation. Nielsen's ten usability heuristics for interface design are the most widely used set in inspection. The heuristics are presented in Table \ref{tab:nielsen-heuristics}.

\begin{table}[h]
\small
\centering
\caption{Nielsen’s 10 Usability Heuristics}
\begin{tabular}{@{}>{\raggedright\arraybackslash}p{0.35\linewidth} >{\raggedright\arraybackslash}p{0.60\linewidth}@{}}
\toprule
\textbf{Heuristic} & \textbf{Description} \\
\midrule
\midrule
1. Visibility of system status & The system should always keep users informed through timely and clear feedback. \\
2. Match between the system and the real world & Speak the users' language and follow real-world conventions. \\
3. User control and freedom & Support undo/redo and easy exits from unintended actions. \\
4. Consistency and standards & Follow platform and industry norms to avoid confusion. \\
5. Error prevention & Prevent problems before they occur with careful design. \\
6. Recognition rather than recall & Minimize memory load with visible elements and clear options. \\
7. Flexibility and efficiency of use & Offer shortcuts and customization for experienced users. \\
8. Aesthetic and minimalist design & Keep interfaces simple, with no irrelevant elements. \\
9. Help users recognize, diagnose, and recover from errors & Use clear messages with constructive guidance. \\
10. Help and documentation & Provide accessible, focused help even when the system is intuitive. \\
\bottomrule
\end{tabular}
\label{tab:nielsen-heuristics}
\end{table}



\subsection{GenAI}
Generative Artificial Intelligence (GenAI) refers to AI models capable of producing novel outputs, such as text, images, code, and interfaces, by learning from large-scale datasets and capturing underlying patterns \cite{10176168}. The emergence of Generative AI and Large Language Models (LLMs) has marked a new era in natural language processing, with LLMs forming a core subset of GenAI specialized in text generation \cite{hagos2024recentadvancesgenerativeai}. Unlike traditional AI, these systems stand out by their creative capacity, enabling tools to generate new artifacts rather than merely predicting or classifying existing ones. 
The creative potential of the new generation of GenAIs has been dramatically changing how practitioners design, develop, and test applications \cite{stige2024artificial}, positioning them at the forefront of current innovation in both Software Engineering (SE) and Human-Computer Interaction (HCI).

In Software Engineering, GenAI is increasingly adopted across the software lifecycle. Tools such as GitHub Copilot and ChatGPT are used for code generation, documentation synthesis, refactoring, and debugging, helping to reduce developer effort and accelerate production by automating repetitive and cognitively demanding tasks \cite{hou2024largelanguagemodelssoftware}. In the Human-Computer Interaction (HCI) and User eXperience (UX) domains, it is being adopted as a co-creative partner that supports designers in ideation, prototyping, and evaluation, such as the automatic generation of personas, design critique, and layout recommendations \cite{stige2024artificial}. They have also been adopted in various activities in the Software Quality Assurance domain, from vulnerability detection to GUI testing \cite{hou2024largelanguagemodelssoftware}. 

Despite their potential benefits, the adoption of GenAI in software development workflows faces several challenges. One of the main challenges is trust and transparency \cite{lu2024ai}. Practitioners must understand how AI produces outputs and whether they can trust them to accept their recommendations. GenAI is known to suffer from ``hallucinations'', i.e., plausible-sounding but incorrect statements, raising concerns regarding the precision and trustworthiness of AI-generated outcomes \cite{zhang2025}. Assessing the quality of such outcomes is also tricky in the context of usability evaluations, given the subjective and evolving nature of UX goals \cite{lu2024ai} and the difficulty these systems have in understanding complex design problems and prototyping ideas \cite{zhou2024exploring}. Context awareness remains another limitation, as AI technologies cannot often understand design intent or contextual nuances, leading to irrelevant or inappropriate suggestions \cite{lu2024ai}. As GenAI becomes more integrated into the software development lifecycle, new methodologies will be required to validate, combine, and contextualize AI outputs.

\subsection{Related Work}
The emergence of LLMs has transformed several computing fields, including usability evaluation. One line of research is using generative AIs to simulate user behavior and generate research data. Yuxuan Lu et al. \cite{lu2025uxagentsimulatingusabilitytesting} introduced UXAGENT, a system that simulates usability testing of web designs using LLM agents to provide early feedback and reduce costs. Similarly, Yoon, Feldt, and Yoo \cite{10638557} introduced DROIDAGENT, an autonomous GUI testing agent for Android applications that uses LLMs to perform intent-driven testing, focusing on high-level, human-like task goals. In the domain of research data generation, H\"{a}m\"{a}l\"{a}inen, Tavast, and Kunnari \cite{10.1145/3544548.3580688} evaluated the potential of the GPT-3 model in generating synthetic user survey data for HCI research. The authors found that GPT-3 can produce credible accounts of HCI experiences, but with some limitations, such as incomplete answers and factual errors.

Another line of research is optimizing user feedback and question formulation through generative AIs. For instance, Eduard et al. \cite{Kuric21112024} explored using GPT-4 to generate follow-up questions in usability studies. Researchers also analyzed the collaboration between conversational AIs and human evaluators. Kuang et al. \cite{10.1145/3613904.3642168} investigated how conversational AI assistants can improve UX evaluation, focusing on the optimal timing of AI automatic suggestions. Complementarily, Yunxing and Jean-Bernard \cite{LIU2024103227} explored using hybrid conversational UIs for the Repertory Grid Technique (RGT) in qualitative research.

There are also studies focused on making comparisons of approaches, which are within the scope of our research. In the context of usability evaluation, comparing different methods is a recurring theme. One of the first studies is from Jeffries et al. \cite{jeffries1991user} in the 90s, who compared heuristic evaluation, software guidelines, cognitive walkthroughs, and usability testing. Since then, several comparisons have been done, mainly between usability inspection and user testing \cite{HVANNBERG2007225,nakamura2020inspect,deKock2009,Hasan01072012}, highlighting that researchers are constantly seeking to identify the most cost-effective approach to improve software quality. In this context, the emergence of generative AIs raised questions on whether such technologies can identify usability defects compared to humans, leading researchers to compare them.

Bisante et al. \cite{10.1145/3656650.3656676} developed CWGPT, a ChatGPT-4-based tool designed to evaluate web interfaces inspired by the ``Cognitive Walkthrough'' method. CWGPT demonstrated high agreement with human experts in identifying usability issues, providing more detailed assessments, and suggestions for improvement. However, sometimes it overlooked visual problems that were evident to human evaluators. In turn, Duan et al. \cite{duan2024} developed a Figma plugin that implements GPT-4 to automate heuristic evaluation and provide feedback on the designed prototypes. The authors evaluated the plugin with experts and concluded that it performs well on poorly designed UIs, but its utility decreases as the UI improves. Meinecke et al. \cite{Meinecke2024} compared the performance of seven inspectors, one expert in usability evaluation, with GPT-4. The authors did not find significant differences in the identification rates of usability issues or severity ratings, concluding that there is a need for more studies with experts.

More recently, Chaves et al. \cite{solariagpt} developed Solaria-GPT, a GPT agent that applies Nielsen's 10 heuristics for analyses of text, images (screenshots), and videos. Their work compared the agent with other models (Qwen 2.5 Max, DeepSeek-V3, Claude 3.7 Sonnet, and Gemini 2.0 Advanced) in terms of accuracy, utility, and new defect rate. However, it lacks comparisons with human experts. In turn, Guerino et al. \cite{guerino2025can} evaluated GPT-4o by analyzing 20 screenshots with a structured prompt based on Nielsen's heuristics, comparing results to experts using defect concordance and severity as metrics.

These works provide valuable contributions to the field, highlighting the potential of generative AIs in usability evaluation. Our work advances this line of research by conducting a systematic and quantitative study on the topic. Unlike previous works, we compared two state-of-the-art generative AIs (Gemini 2.5 Flash and ChatGPT-4o) and a group of experienced usability inspectors to understand the trade-offs between the approaches. Additionally, we analyzed the performance of each approach by calculating established AI evaluation metrics \cite{10.1145/3641289,saleh2025evaluating,duan2024,10.1145/3613904.3642168,10638557}, along with developing a prompt based on standard prompt generation techniques for software development \cite{White2024}. We also investigated the overlap between the problems identified, the characteristics of false-positives, and the strengths and weaknesses of each approach. Finally, we provide implications for practitioners and researchers and devise future research directions, contributing to advancing the field.

\section{Methodology}
In this study, we aimed to compare the performance of GenAIs and human inspectors in identifying usability defects in usability inspection. To scope this study, we followed the Goal-Question-Metric (GQM) paradigm \cite{caldiera1994goal} (see Table \ref{tab:gqm}).

\begin{table}[htbp]
    \centering
    \caption{Goal of the study according to the GQM paradigm.}
    \begin{tabularx}{\columnwidth}{p{2.3cm}|X}
    \toprule
    \textbf{Analyze} & the use of GenAI models (ChatGPT and Gemini) \\
    \textbf{For the purpose of} & evaluating their effectiveness and limitations in usability inspection \\
    \textbf{With respect to} & their ability to identify usability defects compared to human inspectors \\
    \textbf{From the point of view of} & HCI researchers and software quality engineers \\
    \textbf{In the context of} & early-stage usability evaluations of interactive systems based on static interface screenshots. \\
    \bottomrule
    \end{tabularx}
    \label{tab:gqm}
\end{table}

\begin{figure}[htbp]
    \centering
    \includegraphics[width=0.35\textwidth]{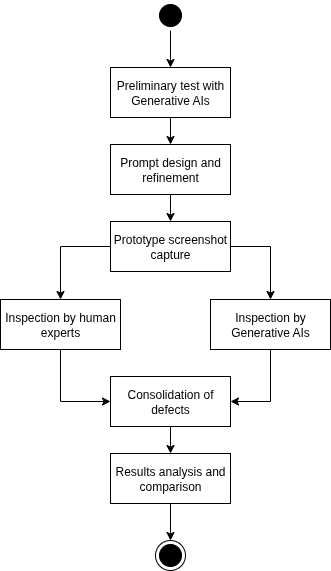}
    \caption{Research methodology.}
    \label{fig:fluxograma}
\end{figure}

Figure \ref{fig:fluxograma} presents our study steps. The following subsections detail each of these steps.

\subsection{Preliminary test with GenAIs}
In this study, we selected two GenAIs: ChatGPT and Gemini. We selected ChatGPT due to its widespread popularity and Gemini due to its growing notoriety. In both cases, we tested the paid versions, specifically GPT-4o and Gemini 2.5 Flash models, which are the standard models.

Initially, we conducted an exploratory test with ChatGPT and Gemini to assess their ability to analyze screenshots of UIs. To do so, we attached a simple login interface. We asked them to inspect it according to Nielsen's 10 heuristics and report any violations of these heuristics, the severity of each violation, and provide improvement suggestions. To avoid the risk of the AIs evaluating interfaces and issues already in their training set, we designed a login screen with usability problems from scratch. As the AIs successfully reported the usability issues, we developed and refined the prompt.

\subsection{Prompt Design and Refinement}
To develop the prompt, we adopted the `Requirements Simulator' pattern in the ``Requirements Elicitation'' category from the work of White et al. \cite{White2024} as the basis (see Table \ref{tab:pattern}. We adapted it to the context of usability inspection by making the AI expect a sequence of images showing the steps of using an application. We also included the name and description of Nielsen's 10 heuristics \footnote{https://www.nngroup.com/articles/ten-usability-heuristics/}, the severity classification, and the desired output format as instructions. The final prompt used in this study can be seen in the Appendix \ref{ap:prompt}.

\begin{table}[ht!]
    \centering
    \caption{Requirements Simulator pattern \cite{White2024}.}
    \begin{tabularx}{\columnwidth}{cX}
        \toprule
         1. & I want you to act as the system. \\
         2. & Use the requirements to guide your behavior. \\
         3. & I will ask you to do X, and you will tell me if X is possible given the requirements. \\
         4. & If X is possible, explain why using the requirements. \\
         5. & If I can't do X based on the requirements, write the missing requirements needed in format Y. \\
         \bottomrule
    \end{tabularx}
    
    \label{tab:pattern}
\end{table}

\subsection{Prototype screenshot capture}
The interactive product selected for inspection was a system prototype developed by two third-year Computer Science students as part of a higher-education institution's Human-Computer Interaction (HCI) course. The system was designed to serve a real client, aiming to assist physicians and nurses in the application of the PEWS (\textbf{P}ediatric \textbf{E}arly \textbf{W}arning \textbf{S}core) protocol in pediatric intensive care units (ICUs). PEWS is a protocol designed to consistently assess pediatric patients' conditions using physiological indicators, such as cardiovascular status, respiratory status, nebulizer use, and persistent post-surgical emesis \cite{gold2014evaluating}.

The client selected this prototype among five others as the most suitable for her needs. For analysis, we captured screenshots of each prototype screen, including interactions with menus and forms, in a sequential order (16 screenshots in total).

\subsection{GenAIs' Inspection}
After refining the prompt, we started the inspection process. One of the authors of this work worked with Gemini, and another with ChatGPT. Each author started the inspection by prompting the AI. Next, they uploaded each screenshot, one at a time. The screenshots were organized and uploaded to the AIs in a logical sequence to simulate a real usage of the system. After submitting all the screenshots, they organized the AIs' outputs in a Google Sheets spreadsheet.

\subsection{Experts' Inspection}
In this study, we selected four usability inspection experts by convenience. The inspectors group comprised a requirements analyst, a software analyst, a product owner, and a Computer Science professor—all males, with an average age of 31.25. Two inspectors have PhD degrees, and two are PhD students with master's degrees, all in Computer Science. All inspectors had extensive experience with usability inspection. Among them, three had already conducted inspections in several projects, though not regularly, while one performed inspections constantly as part of his work. 

Initially, each inspector filled a characterization questionnaire and an informed consent form (ICF) informing them of the study's objective, voluntary participation, and the processing of the collected data. Each participant received, by e-mail, Nielsen's set of heuristics and a spreadsheet from Google Sheets to report the problems identified, comprising the following fields: inspector's name, inspection start/end time, problem location, violated heuristic(s), problem description, severity, and improvement suggestions. Each inspector performed the inspection individually and asynchronously, without the use of AI.



\subsection{Consolidation of defects}
To consolidate the defects, one of the authors of this work created a spreadsheet and aggregated the participants' and AIs' reports. He assigned a unique ID for each item of the reports (i.e., the discrepancies). Then, he read each of the discrepancies and classified them as follows:

\begin{itemize}
    \item \textbf{False-positive:} discrepancy that did not represent a real problem (e.g., functionalities that are out of scope of the software, opinions, suggestions, suppositions).
    \item \textbf{Defect:} validated discrepancy considered as defect.
    \item \textbf{Duplicate:} defect already reported.
\end{itemize}

For each duplicated defect, he assigned the ID of the similar defect. After finishing the analysis, the third author, an HCI and usability inspection expert, reviewed the spreadsheet by including comments on the items with diverging opinions. Next, the second author, also an HCI and usability inspection expert, reviewed the spreadsheet and commented. Diverging items were discussed until a consensus was reached. The spreadsheet with all discrepancies and analyses is available at Figshare \footnote{https://doi.org/10.6084/m9.figshare.29631536}.

\subsection{Evaluation metrics}
\label{subsec:metrics}
We adopted a set of metrics to evaluate the performance of both human inspectors and GenAIs. We calculated effectiveness and efficiency for within-group performance (i.e., among inspectors or GenAIs), as used in a previous paper by two of the authors \cite{nakamura2020inspect}. We computed \textbf{effectiveness} as the number of defects identified by inspector/AI $i$ by the number of unique defects from the group $j$ which $i$ belongs to (i.e., either the inspector group or the AI group), as follows:
\begin{equation}
\label{form:effectiveness}
    Effectiveness_{i,j} = \frac{Defects_{i}}{UniqueDefects_{j}}
\end{equation}

To compute \textbf{efficiency}, we divided the number of defects identified by $i$ by the amount of time $i$ spent performing the inspection:
\begin{equation}
    \label{form:efficiency}
    Efficiency_{i} = \frac{Defects_{i}}{time_{i}}
\end{equation}

For between-group performance comparisons (inspectors, Gemini, and ChatGPT), we chose precision, recall, and F1-score as evaluation metrics. These were selected based on studies validating GenAI model performance \cite{10.1145/3641289,saleh2025evaluating} and applied to usability defects \cite{duan2024,10.1145/3613904.3642168,10638557}. To calculate \textbf{precision}, we divided the number of true positives (TP), where TP represents defects identified by the group, by the total number of discrepancies reported by that group, which includes the TP and false positives (FP):
 

\begin{equation}
    \label{form:precision}
    Precision = \frac{TP}{TP+FP}
\end{equation}

To calculate \textbf{recall}, we divided the number of unique defects identified by the group (TP) by the number of all unique defects identified in the study (TP+FN):
\begin{equation}
    \label{form:recall}
    Recall = \frac{TP}{TP+FN}
\end{equation}

Finally, we calculated the \textbf{F1-score} as follows:
\begin{equation}
    \label{form:f1}
    F1 = \frac{2*Precision*Recall}{Precision+Recall}
\end{equation}

To identify whether the differences between the groups are significant, we applied Fisher's exact test \cite{fisher1970statistical}. It is a statistical test that verifies the significance of the deviation from the null hypothesis between two elements \cite{SILVEIRA2022111223}. Despite being applicable to any sample size, it is frequently adopted in studies with less than 30 samples, having been applied in previous usability evaluation studies \cite{Meinecke2024, SILVEIRA2022111223, Britto2009,cockton2001understanding}.

\section{Results}
In this section, we present the results of the inspection. We divided the results into subsections to facilitate their understanding.

\subsection{Number of defects overview}

Table \ref{tab:overview-inspectors} presents an overview of the results from inspectors. A total of 108 discrepancies were reported, ranging from 17 to 37 per inspector (average 27). Among them, nine were considered false-positives, such as the following:

\begin{quote} 
    ``\textit{[82] I would like to be able to search by reviewer}''
\end{quote}
\begin{quote} 
    ``\textit{[66] There is no option to export or print historical data for clinical follow-up.}''
\end{quote}
\begin{quote} 
    ``\textit{[89] The purpose of this screen was not clear.}''
\end{quote}

\begin{table}[h]
\caption{Performance of the inspectors group.}
\label{tab:overview-inspectors}
\begin{tabularx}{\columnwidth}{lcccc}
    \toprule
     & I1 & I2 & I3 & I4 \\ 
     \midrule
    Discrepancies & 17 & 19 & 37 & 35 \\ 
    False-Positives (FP) & 1 (5.88\%) & 0 (0\%) & 3 (8.11\%) & 5 (14.29\%) \\ 
    Defects (TP) & 16 & 19 & 34 & 30 \\ 
    Duplicates & 0 & 7 & 12 & 8 \\
    Unique defects (TP) & 16 & 12 & 22 & 22 \\ 
    Time (minutes) & 35 & 45 & 95 & 170 \\ 
    Effectiveness & 22.22\% & 26.39\% & 47.22\% & 41.67\% \\ 
    Efficiency & 27.43 & 25.33 & 21.47 & 10.59 \\ 
    \midrule
    Avg. Defects & \multicolumn{4}{c}{24.75} \\ 
    \midrule
    Avg. Effectiveness & \multicolumn{4}{c}{34.38\%} \\ 
    \midrule
    Avg. Efficiency & \multicolumn{4}{c}{21.21} \\
    \bottomrule
   \end{tabularx}
\end{table}

These false-positives do not address a defect, given that they are related to inspectors' preferences (discrepancy 82), issues that are out of the scope of the prototype (66), and generic reports (89). In the end, the number of defects ranged from 16 to 34 (average 24.75), which resulted in an effectiveness of 34.38\% on average.

The `duplicates' row highlights the level of redundancy among inspectors. From the initial set identified by I1, each new inspector brought another set of unique defects, but also addressed issues covered previously by other inspectors. This result suggests that inspectors focus on different aspects of the interface, reinforcing the need for different inspectors to cover a larger set of problems.

The time spent by each inspector varied from 35 to 170 minutes. Due to time constraints, participant I4 performed the inspection in three rounds on different days, resulting in a greater time than the others who conducted the inspection all at once. On average, inspectors had an efficiency of 21.21 defects per hour.

Regarding the GenAIs (see Table \ref{tab:overview-ais}), ChatGPT reported a greater number of discrepancies (100) than Gemini (76); both reported more discrepancies than inspectors individually. In turn, the AIs produced more false-positives than inspectors (13 from Gemini and 18 from ChatGPT). Notably, the reasons for the false-positives varied for each AI.

\begin{table}[h!]
\caption{Performance of the GenAIs.}
\label{tab:overview-ais}
\begin{tabular}{lcc}
    \toprule
    & Gemini & ChatGPT \\ 
    \midrule
    Discrepancies & 76 & 100 \\ 
    False-Positives (FP) & 13 (17.11\%) & 18 (18\%) \\ 
    Defects (TP) & 63 & 82 \\ 
    Duplicates & 8 & 30 \\ 
    Unique defects (TP) & 55 & 52 \\
    Shared defects & \multicolumn{2}{c}{18} \\ 
    \midrule
    Effectiveness & 61.80\% & 58.43\% \\ 
    \midrule
    Avg. Defects & \multicolumn{2}{c}{72.5} \\ 
    \midrule
    Avg. Effectiveness & \multicolumn{2}{c}{60.11\%} \\
    \bottomrule
   \end{tabular}
\end{table}

Regarding Gemini, the false-positives were spread across different types, from a supposed lack of feedback to visual aesthetics:

\begin{quote} 
    ``\textit{[UI-006] The `Child's Name' search bar doesn't provide immediate feedback after starting a search. While results can be updated, a visual indicator that the search is in progress (e.g., a spinning loading icon) would improve visibility of the system status.}''
\end{quote}
\begin{quote} 
    ``\textit{[UI-048] The label placement and alignment are slightly inconsistent. The labels `Patient Name' and `Age Group' are above their fields, while `Bed,' `Diagnosis,' `DIH,' and `Evaluation Date' have their labels to the left.}''
\end{quote}
\begin{quote}
    ``\textit{[UI-072] The `Back' and `Home' buttons use different shades of green, even though both are primary action buttons in this context.}''
\end{quote}
\begin{quote} 
    ``\textit{[UI-061] The input fields for `Respiratory Rate' and `Heart Rate' are basic text inputs. There are no numeric input restrictions, minimum/maximum values, or real-time validation to prevent illogical input (e.g., negative numbers, excessively large numbers).}''
\end{quote}

The discrepancy UI-006 reports a lack of feedback that cannot be inferred from a screenshot. UI-048 and UI-072, in turn, highlight problems in the alignment of the items and colors. However, all labels are on the top of the input fields, while the color of the buttons is the same (\#006e42). Finally, UI-061 reports a lack of validation that cannot be assessed through a screenshot.

Regarding ChatGPT, it tended to focus too much on system feedback. Seven out of 18 were related to the lack of feedback after acting, which cannot be inferred from a screenshot. There were also some false-positives related to visual inconsistencies and input validation, as in the quotations below:

\begin{quote}
    ``\textit{[UI-143] The visual alignment between toggle switches and corresponding labels is inconsistent, potentially causing confusion about what each toggle controls.}''
\end{quote}
\begin{quote} 
    ``\textit{[UI-133] There is no validation ensuring that  `DIH' (Days Admitted to Date [`Dias Internados até Hoje' in portuguese]) does not have a future date compared to `Assessment Date'.}''
\end{quote}

\begin{figure}
    \centering
    \includegraphics[width=1\linewidth]{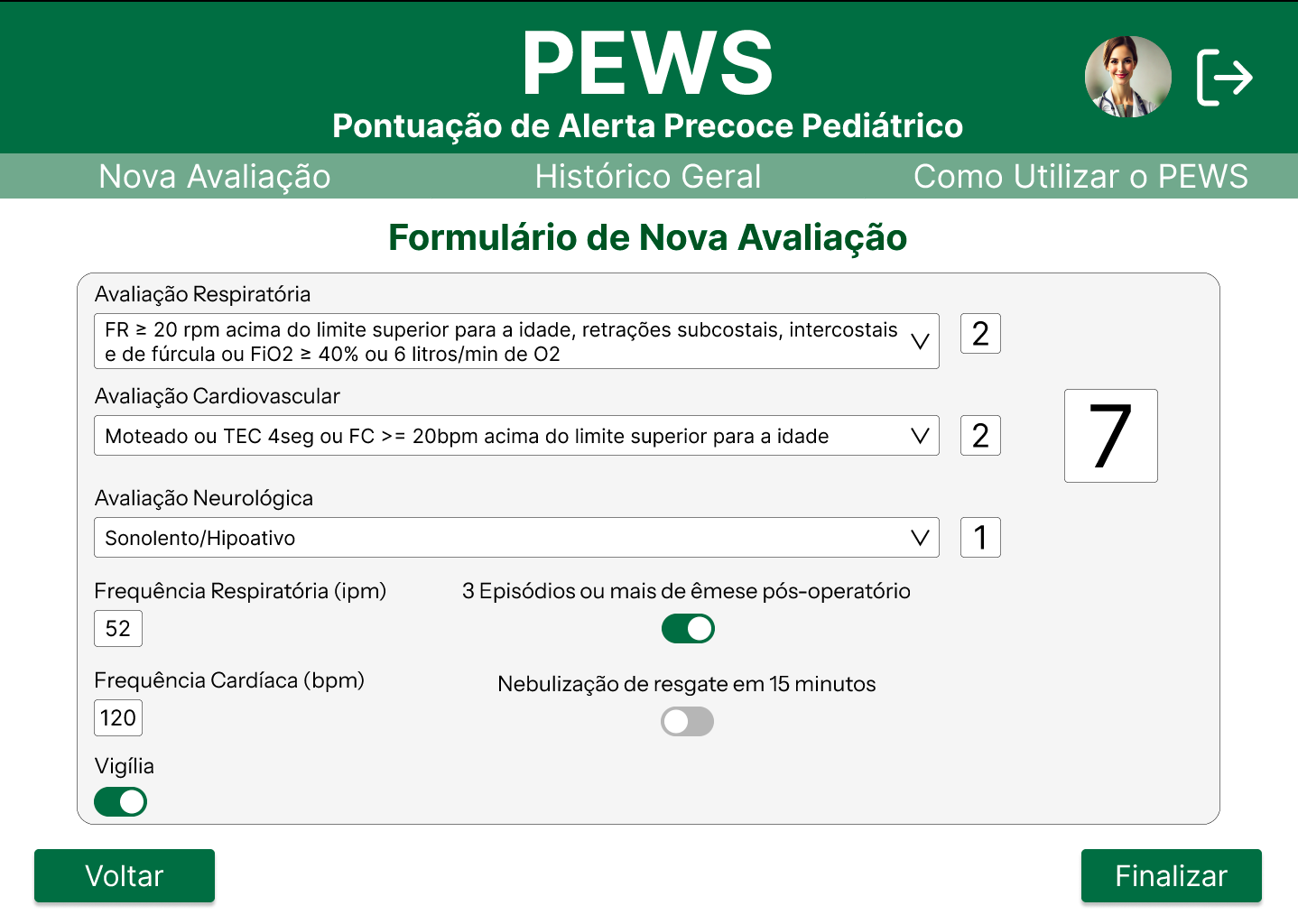}
    \caption{Patient's clinical assessment form.}
    \label{fig:prototype}
\end{figure}

Consider Figure \ref{fig:prototype}, which presents a filled PEWS form. Discrepancy UI-143 reports a problem in the visual alignment of the toggles, but all are consistently below their labels. UI-133, in turn, cannot be assessed from the screenshot, as this specific scenario was not presented in the screenshots. 

In general, both AIs made assumptions that cannot be verified from the screenshots, something that did not happen with human inspectors. In turn, inspectors sometimes were influenced by their personal opinions, providing suggestions and reporting issues that were not real problems. 

ChatGPT reported a greater number (30) than Gemini (8) regarding duplicated defects. Most of the duplicated defects from the AIs were because they reported the same defect using another heuristic. While inspectors assigned more than one heuristic to a discrepancy in the inspection form when applicable, the AIs generated different discrepancies to report the same problem, but with a different heuristic. ChatGPT also tended to report the same defect in screenshots representing the user interaction with a given functionality (e.g., the form is empty in one screenshot, while in the following screenshot, all the fields are filled). In the end, the effectiveness of the Gemini and ChatGPT was about 60\%, indicating that they covered most of the defects identified by the AI group.

\subsection{Defects distribution}

Table \ref{tab:defects-by-group} presents the total of unique defects (defects minus duplicates) and the number of new defects by group. From the initial set of 72 defects identified by inspectors, Gemini identified another set of 35 new defects, while ChatGPT contributed another 20 novel ones, resulting in a total of 127 unique defects. Regarding the number of defects, human inspectors obtained the best performance by identifying 72 out of 127 defects, followed by Gemini (55) and ChatGPT (52). 

To analyze the level of coverage between the groups, we identified overlapping defects (see Table \ref{tab:defects-overlap}). Gemini identified around a fourth of the defects reported by inspectors (20 out of 72), while ChatGPT identified a third of the defects from inspectors (24 out of 72). Together, both AIs addressed almost half of the defects reported by inspectors (33 out of 72). 

\begin{table}
    \caption{Number of defects by group.}
    \label{tab:defects-by-group}
    \begin{tabular}{lccc}
        \toprule
        & Inspectors & Gemini & ChatGPT \\
        \midrule
        Unique & 72 & 55 & 52 \\
        New defects & 72 & 35 & 20 \\
        \midrule
        Total defects & \multicolumn{3}{c}{127}  \\
        \bottomrule
    \end{tabular}
\end{table}

\begin{table*}[bp]
    \caption{Defects overlap among groups.}
    \label{tab:defects-overlap}
    \begin{tabular}{lccc}
        \toprule
        & $Gemini \cap  Inspectors$ & $ChatGPT \cap Inspectors$ & $(Gemini \cup ChatGPT)  \cap  Inspectors$ \\
        \midrule
        Defects & 20 / 72 & 24 / 72 & 33 / 72 \\ 
        Percentage & 27.78\% & 33.33\% & 45.83\% \\
        \bottomrule
    \end{tabular}
\end{table*}

\begin{table*}[bp]
  \caption{Performance between inspectors, GenAIs, and their combinations.}
  \label{tab:precision-recall-f1}
  \begin{tabular}{lcccccc}
    \toprule
    & Inspectors & Gemini & ChatGPT & $Gemini \cup ChatGPT$ & $Inspectors \cup Gemini$ & $Inspectors \cup ChatGPT$ \\
    \midrule
    Precision & 0.917 & 0.829 & 0.820 & 0.824 & 0.880 & 0.870 \\
    Recall & 0.567 & 0.433 & 0.409 & 0.701 & 0.843 & 0.724 \\
    F1-score & 0.701 & 0.569 & 0.546 & 0.757 & 0.861 & 0.791 \\
    \bottomrule
  \end{tabular}
\end{table*}

\subsection{Precision, Recall, and F1-score}
In Table \ref{tab:precision-recall-f1}, we present the precision, recall, and F1-score metrics results by group and combination of groups. Regarding precision, the inspectors group obtained the best performance (.917), addressing many defects with only a few false-positives. The GenAIs achieved similar performance, with Gemini (.829) slightly outperforming ChatGPT (.820). Fisher's exact test indicated that the difference in precision between inspectors and ChatGPT is significant (p = .042), but not with Gemini (p = .105). The combination of the groups did not result in better precision.

The results for recall revealed that only the inspector group identified the majority of the defects (.567). Gemini performed slightly better (.433) than ChatGPT (.409). Fisher's exact test revealed that inspectors covered significantly more problems than ChatGPT (p = .017) and Gemini (p = .044). The combination of groups improved recall due to the complementarity of the defects identified. The best combination was between Gemini and Inspectors (.843). 

Finally, regarding F1-score, inspectors achieved the best individual performance (.701), followed by Gemini (.569) and ChatGPT (.546). The combination of inspectors and Gemini produced the best results, with a good balance of precision and recall (.861). Combining ChatGPT and Gemini resulted in a better F1-score than inspectors, with a good coverage of the defects at the cost of precision.

\section{Discussion}
In this section, we discuss the main findings of this study. To facilitate the understanding of the results, we divided them into subsections.

\subsection{Effectiveness and Efficiency}
The inspectors group identified 72 unique defects, the highest among all groups. Such results highlight that human inspectors can contextualize issues even with limited information, i.e., static screenshots, and recognize problems that emerge from human-centered reasoning and interface interpretation. However, their performance varied widely across individuals. Their effectiveness, for instance, varied from 22.22\% to 47.22\%, highlighting that each inspector focuses on different aspects of the UI, which reinforces the need for different inspectors to identify a broader range of usability problems, as already identified in previous usability inspection studies \cite{nielsen1995technology,costabile2002guidelines}. 

The efficiency of inspectors also varied significantly, ranging from 10.59 to 27.43 defects/hour. The performance of inspectors may be influenced by different factors, such as the characteristics inherent to the inspector (some may seek to inspect the UI thoroughly, while others may pass through), experience, time commitment, and possibly cognitive fatigue. Such characteristics may reduce the applicability of human inspection in time-constrained projects. Although we did not calculate efficiency, given that such a comparison with human inspectors would not be fair, GenAIs can analyze UI screenshots in a few seconds, which is a strong practical advantage.

Regarding the GenAIs, both Gemini and ChatGPT identified more defects than any of the inspectors individually, resulting in greater individual effectiveness (61.8\% and 58.43\% respectively). Gemini slightly outperformed ChatGPT in effectiveness, but both could identify defects not recognized by inspectors. However, while Meinecke et al. \cite{Meinecke2024} found that there was no significant difference between AI and inspectors, our results indicate that the AIs reported fewer defects (54 from Gemini and 52 from ChatGPT) compared to the group of inspectors (72). Such a difference may be related to the profile of inspectors. While in our study, all four inspectors were experts with extensive experience with usability inspection, only one out of seven inspectors in the study of Meinecke et al. \cite{Meinecke2024} was an expert. Such results indicate that expert human inspectors still bring more diversity in findings than AIs.

\subsection{False-positives}
Inspectors had a lower false-positive rate (ranging from 0 to 14.29\%) than Gemini (17.11\%) and ChatGPT (18\%), a finding that corroborates with Meinecke et al. \cite{Meinecke2024}. In general, human judgment was more grounded in what could be verified from the screenshots, without speculating about possible defects and interaction failures. However, in some cases, inspectors were driven by personal preferences, e.g., features that were desirable but not required for task execution, leading sometimes to out-of-scope comments in the problem reporting form. Such results point to human bias, which can be strengths (creativity, context awareness) and weaknesses (subjectivity).

False-positives from AIs, in turn, consisted mainly of violations of visual aesthetics standards and assumptions beyond what could be determined from screenshots (e.g., feedback, validations). Regarding the first reason, the AIs overemphasized visual details, such as label alignment and color shades, which in most cases were untrue. Gemini, for example, reported alignment issues that did not exist and misinterpreted aesthetic aspects. Regarding the second, ChatGPT emphasized missing system feedback and input constraints too much. This highlights that GenAIs can infer behavior based on visual clues without needing screenshots for every user interaction. On the one hand, such feedback may be valuable for practitioners in improving product quality when moving from the design phase to implementation. On the other hand, it may result in false-positives that can overwhelm practitioners with unnecessary details in the early stages of the project. These findings also open research possibilities to investigate whether GenAIs can evaluate the usability from user interaction videos.

In summary, while both groups generated some erroneous reports, AIs tended to hallucinate functionality and aesthetics, whereas humans tended to make assumptions based on their expectations and preferences. These findings suggest different types of interpretative bias across evaluation agents.

\subsection{Duplicated Defects}
The duplication of defects was a notable issue for ChatGPT, which reported 30 duplicates, while Gemini reported only 8. AIs duplicated problems due to two main reasons: (1) they reported the same defect in the same screenshot using different heuristics (mainly with ChatGPT), (2) they identified the same problem in different screenshots (especially between screenshots representing user interaction, such as when one shows an empty state and another a filled state). Inspectors, in contrast, used a single discrepancy and associated multiple heuristics. Moreover, they did not repeat the problem, even in the second situation faced by the AIs, showing more efficient grouping of findings. Such results indicate differences in how each AI accesses its memory, performs pattern grouping, or follows instructions. 

\subsection{Precision, Recall, and F1-score}
Overall, inspectors performed better than AIs, reporting fewer false-positives (precision = .917) and identifying most of the problems in the study (recall = .567). However, our study revealed that combining approaches may result in a more comprehensive evaluation. For instance, combining Gemini and ChatGPT outcomes resulted in a good coverage of defects (recall = .701). However, the combination that produced the best results was between inspectors and AIs, particularly with Gemini (precision = .880, recall = .843, f1-score = .861), highlighting the potential benefit of hybrid inspection approaches.

The analysis on defect coverage revealed low redundancy across groups, indicating that each group identified a unique set of defects. ChatGPT overlapped more with inspectors than Gemini (33.3\% vs 27.78\%), but neither AI nor their combination (45.83\%) could cover most of the problems identified by inspectors. In turn, it is noteworthy that both contributed with 55 defects not seen by any inspector (an increase of 76.39\% over the initial set). Such findings suggest that the AIs cannot fully replace the human perspective, but can add value as complementary agents to identify a broader set of defects.

\subsection{Will AI also replace inspectors?}
Answering the question in our paper's title, our study's results suggest that AI, in its current stage, cannot fully replace human inspectors. While GenAIs can identify a significant number of issues in a few times, they often make unverifiable assumptions (e.g., lack of feedback, input validation, behavior), while emphasizing aesthetic inconsistencies. Thus, their false-positive rates were higher than those of human inspectors. Moreover, they failed to detect many usability problems that humans had identified. Such results highlight that domain knowledge, contextual reasoning, and user-centered interpretation of the interface inherent to humans differentiate GenAIs and experts. Human inspectors, in turn, achieved higher precision, reported more defects, and produced less duplication. However, such results come at the cost of time and variability in performance.

Collaboration seems to be the key to achieving a balance between depth, breadth, and efficiency in usability inspections. Rather than viewing AIs as replacements, the evidence supports their role as augmentation tools that can increase efficiency, coverage, and scalability when integrated into the usability inspection process. Therefore, AI will not replace human inspectors but reshape how inspections are conducted, leading to several implications for the software quality field.

\begin{tcolorbox}[colback=white!100!white,
                  colframe=black!100!black,
                  title=Key Takeaways,
                  fonttitle=\bfseries,
                  coltitle=white,
                  boxsep=3pt,
                  left=3pt,
                  right=3pt,
                  top=3pt,
                  bottom=3pt,
                  sharp corners]
\begin{itemize}
    \item Human inspectors reported more unique defects with higher precision and less duplication.
    \item Inspectors sometimes included subjective or out-of-scope comments, but they were generally grounded in evidence.
    \item GenAIs were faster and more effective individually, but produced more false-positives and redundancies.
\end{itemize}

\end{tcolorbox}

\begin{tcolorbox}[colback=white!100!white,
                  colframe=black!100!black,
                  fonttitle=\bfseries,
                  coltitle=white,
                  boxsep=3pt,
                  left=3pt,
                  right=3pt,
                  top=3pt,
                  bottom=3pt,
                  sharp corners]
\begin{itemize}
    \item GenAIs made unverifiable assumptions and overemphasized visual aesthetics.
    \item The combination of humans and AIs resulted in the best trade-off between precision and defect coverage.
\end{itemize}

\end{tcolorbox}

\subsection{Implications for Software Quality}
Our results have several implications for the software quality domain. First, practitioners should rethink usability evaluation practices. While the experience and perception of human inspectors remain essential, the traditional model of manual heuristic evaluation may benefit from introducing GenAIs into the pipeline.

GenAIs may be particularly valuable within time-constrained contexts, where they could provide an initial layer of inspections, which human experts can further validate, contextualize, and expand their findings. The results from the AI could be used to screen the UI, flag screens with a high volume of potential defects, and allow inspectors to prioritize those that require a more in-depth review.

They can also be helpful in the design review at early stages of software development, when designers are still iterating on low-fidelity wireframes. By doing so, designers could quickly identify inconsistencies, missing labels, and nonconformance with standards. At this stage, the cost of change is low, and the feedback of the AIs, even if imperfect, can help inform rapid iterations.

The differences in defect types suggest that Quality Assurance (QA) teams can perform screening of the issues reported by AIs, using confidence scores or filters to prioritize reports. With AIs taking on preliminary review roles, usability teams may scale their evaluations across multiple products more efficiently, ensuring higher coverage without linear cost increases.

Another possibility is integrating GenAIs into Continuous Integration and Continuous Delivery (CI/CD) pipelines in later stages of the project, when the UI is established. For instance, after a front-end update, the AI could detect deviations from patterns and usability heuristics, allowing practitioners to identify the issue and make fixes.

Considering the use of AI in software development practices, educational programs in Human-Computer Interaction and Software Engineering may also need to adapt their curricula to cover this integration while addressing topics related to training, interpreting, and refining AI-generated feedback. Students should learn about the potential of AIs and their limitations to be able to critically analyze the AI outputs, identify false-positives, and recognize when they trespass the limits of available data by making assumptions.

\section{Threats to Validity}

Following the classification proposed by Wohlin et al. \cite{wohlin2012experimentation}, we discuss below the main threats to the validity of our study.

\textbf{Construct validity} may be affected by using prompts, which can influence the behavior and output of generative AI models. To mitigate this threat, we iteratively refined the prompt based on established guidelines and patterns from the literature \cite{White2024}. 

\textbf{Internal validity} concerns arise from the number and profile of inspectors, which may influence the types of usability issues identified. Although we did not conduct evaluations until full saturation was reached, i.e., the point at which new evaluators contribute few or no additional problems, we adopted Nielsen’s recommendation of using three to five evaluators \cite{nielsen1994heuristic}. This range has been shown to uncover approximately 85\% of usability problems in typical inspection scenarios and is widely accepted in the literature.

\textbf{External validity} is potentially limited by the inherent variability in outputs generated by large language models. Even with the same prompt, generative AIs may produce slightly different results across executions. Another point is that the inspection was limited to one system, which limits generalizability to other domains and application types. Therefore, our findings should be interpreted within the specific context described in our study, including the AI models, configurations, and procedures used. While these limit the generalizability of the results, they strengthen their applicability within similar settings.

\textbf{Conclusion validity} may be threatened by potential errors in statistical interpretation or the inappropriate use of analytical techniques. To mitigate this risk, independent researchers carefully reviewed all statistical analyses in pairs. Additionally, we employed multiple complementary metrics to support our inferences, which increases the robustness and reliability of the conclusions drawn from the data.



\section{Conclusion}




In this study, we explored the performance of GenAI models in supporting the usability inspection process. Usability inspection traditionally demands considerable time and specialized expertise from the inspectors. We conducted a comparative analysis involving a software prototype evaluated by four inspectors and two GenAI models, ChatGPT (GPT-4o) and Gemini (2.5 Flash). By employing metrics such as precision, recall, and F1-score, we assessed the strengths and limitations of each approach, aiming to understand how they differ and how they might complement each other.

Our results show that inspectors were more effective in identifying a broader range of usability issues, with higher precision and fewer duplicated findings. Their ability to interpret context and recognize nuanced interface problems proved to be a distinguishing factor. On the other hand, GenAI models demonstrated a lower time to perform tasks and greater consistency, producing results in a fraction of the time required by humans. Despite this efficiency, both GenAI models exhibited low recall and a higher rate of false positives, particularly when interpreting visual or aesthetic elements. ChatGPT reported more defects overall, while Gemini achieved slightly better precision. When we combined the findings from both GenAIs, the coverage improved, although it still did not match the diversity and depth of human inspection. These findings reinforce that while GenAIs offer valuable support, particularly in the early evaluation phases, they are not yet suitable replacements for expert inspectors.    

For future work, we plan to replicate this study using additional GenAI models, including free-access versions, to evaluate whether similar performance patterns hold under different configurations and accessibility conditions. We also intend to include novice inspectors and incorporate severity classification in the analysis, allowing us to examine whether humans or AIs are more effective in detecting high-impact usability problems. Additionally, we plan to investigate the applicability of AIs in other steps of usability inspection, such as defect disambiguation, which requires a significant amount of time when performed manually. These directions will contribute to a more comprehensive understanding of the practical role that generative AIs can play in usability inspection processes and how they can be integrated meaningfully into the software quality process. 



\bibliographystyle{ACM-Reference-Format}
\bibliography{main}

\appendix

\section{Final prompt}
I want you to act as a Human-Computer Interaction expert in User eXperience and usability inspection. Use the following 10 Nielsen's heuristics to guide your behavior. Next, I will provide screenshots, one at a time, showing the steps of using an application. You will inspect the screenshots based on the heuristics. Report all distinct violations for each heuristic, even if they are cosmetic. Do not omit minor issues. For each violation, please describe how the problem violates the heuristic, define its severity, and suggest how to fix it.\\

Severity rating: \\
1 = Cosmetic issue (fix only if time permits) \\
2 = Minor usability problem (low priority) \\
3 = Major usability problem (high priority) \\
4 = Usability catastrophe (must fix before release)\\

Respond ONLY in the following JSON format: \\
\begin{verbatim}
[
  {
    "heuristic": str,
    "violations": [
      {
        "description": str,
        "severity": int,
        "solution": str
      }
    ]
  }
]
\end{verbatim}

Heuristic name: H1- Visibility of System Status \\
Heuristic definition: The design should always keep users informed about what is going on, through appropriate feedback within a reasonable amount of time.\\

Heuristic name: H2- Match Between the System and the Real World\\
Heuristic definition: The design should speak the users' language. Use words, phrases, and concepts familiar to the user, rather than internal jargon. Follow real-world conventions, making information appear in a natural and logical order.\\

Heuristic name: H3- User Control and Freedom\\
Heuristic definition: Users often perform actions by mistake. They need a clearly marked "emergency exit" to leave the unwanted action without having to go through an extended process.\\

Heuristic name: H4- Consistency and Standards\\
Heuristic definition: Users should not have to wonder whether different words, situations, or actions mean the same thing. Follow platform and industry conventions.\\

Heuristic name: H5- Error Prevention\\
Heuristic definition: Good error messages are important, but the best designs carefully prevent problems from occurring in the first place. Either eliminate error-prone conditions, or check for them and present users with a confirmation option before they commit to the action.\\

Heuristic name: H6- Recognition Rather than Recall\\
Heuristic definition: Minimize the user's memory load by making elements, actions, and options visible. The user should not have to remember information from one part of the interface to another. Information required to use the design (e.g. field labels or menu items) should be visible or easily retrievable when needed.\\

Heuristic name: H7- Flexibility and Efficiency of Use\\
Heuristic definition: Shortcuts — hidden from novice users — may speed up the interaction for the expert user so that the design can cater to both inexperienced and experienced users. Allow users to tailor frequent actions.\\

Heuristic name: H8- Aesthetic and Minimalist Design\\
Heuristic definition: Interfaces should not contain information that is irrelevant or rarely needed. Every extra unit of information in an interface competes with the relevant units of information and diminishes their relative visibility.\\

Heuristic name: H9- Help Users Recognize, Diagnose, and Recover from Errors\\
Heuristic definition: Error messages should be expressed in plain language (no error codes), precisely indicate the problem, and constructively suggest a solution.\\

Heuristic name: H10- Help and Documentation \\
Heuristic definition: It’s best if the system doesn’t need any additional explanation. However, it may be necessary to provide documentation to help users understand how to complete their tasks.
\label{ap:prompt}

\end{document}